\documentstyle[12pt]{article}

\makeatletter                                                           

\def\section{\@startsection {section}{1}{\z@}{-1.5ex plus -.5ex
minus -.2ex}{1ex plus .2ex}{\large\bf}}                                 

\def\@thmcountersep{}                                                   

\long\def\@makecaption#1#2{\vskip 10pt
\setbox\@tempboxa\hbox{#1. #2}   
   \ifdim \wd\@tempboxa >\hsize   
       #1. #2\par                 
     \else                        
       \hbox to\hsize{\hfil\box\@tempboxa\hfil}                         
   \fi}                                                                 

\begin{document}

\title{No Time Machines from Lightlike Sources \\ in 2+1 Gravity}

\author{
{\em S. Deser}~\thanks{The Martin Fisher School of Physics, Brandeis
University, Waltham, MA 02254, USA.  }
\and
{\em Alan R. Steif}~$^*$\thanks{Present address: DAMTP, Cambridge
University, Silver St., Cambridge, CB3 9EW, U.K. ~~~~~~~~~
E-MAIL: deser@brandeis; ars1001@amtp.cam.ac.uk.}}

\date{} 

\flushbottom                                                           

\maketitle
\vspace{-10pt} 

\begin{abstract}
We extend the argument that spacetimes generated by two timelike
particles in $D$=3 gravity (or equivalently by parallel-moving cosmic
strings in
$D$=4) permit closed timelike curves (CTC) only at the price of Misner
identifications that correspond to unphysical boundary conditions at
spatial infinity
and to a tachyonic center of mass. Here we analyze geometries one or
both of
whose sources are lightlike.  We make manifest both the presence of
CTC at
spatial infinity if they are present at all, and the tachyonic character of
the
system: As the total energy surpasses its tachyonic bound, CTC first
begin to
form at spatial infinity, then spread to the interior as the energy
increases further. We then show that, in contrast, CTC are entirely
forbidden in topologically massive gravity for geometries generated by
lightlike sources.
\end{abstract}

\def\TH{\Theta}
\def\D{\Delta}
\def\z{\zeta}
\def\L{\Lambda}
\def\x{\chi}
\def\P{\Pi}
\def\s{\sigma} \def\S{\Sigma}
\def\t{\tau}
\def\y{\upsilon} \def\Y{\Upsilon}
\def\o{\omega}
\def\th{\theta}
\def\g{\gamma} \def\G{\Gamma}
\def\a{\alpha}
\def\b{\beta}
\def\d{\delta} \def\D{\Delta}
\def\e{\epsilon}
\def\z{\zeta}
\def\k{\kappa}
\def\l{\lambda} \def\L{\Lambda}
\def\m{\mu}
\def\n{\nu}
\def\x{\chi}
\def\p{\pi} \def\P{\Pi}
\def\r{\rho}
\def\s{\sigma} \def\S{\Sigma}
\def\t{\tau}
\def\y{\upsilon} \def\Y{\Upsilon}
\def\o{\omega} \def\O{\Omega}
\def\cos{{\rm cos}\;}
\def\sin{{\rm sin}\;}
\def\sn{{\rm sin}\;{m\over 2}}
\def\cs{{\rm cos }\;{m\over 2} }
\def\tn{{\rm tan}\; {m\over 2}}
\def\pa{\partial}
\def\ra{\rightarrow}
\def\pr{\prime}
\def\xvec{{\bf x}}
\def\xvecpr{{\bf x^{\pr}}}

Among the many fundamental contributions by Charlie Misner to
general relativity is his study of pathologies of Einstein geometries,
particularly NUT spaces, which in his words are ``counterexamples to
almost everything"; in particular they can possess closed timelike
curves (CTC).  As with other farsighted
results of his which were only
appreciated later, this 25-year old one finds a resonance in very recent
studies of conditions under which CTC can appear in
apparently physical settings, but in fact require unphysical boundary
conditions engendered by identifications very similar to those he
discovered.  In this paper,
dedicated to him on his 60th birthday, we review and extend some of
this current work. We hope it brings back pleasant memories.

{\it To be published
in a Festschrift for C.W. Misner, Cambridge University Press.}

\hspace*{5in}BRX TH--336

\section{Introduction} 

Originally constructed by G\"{o}del \cite{godel1}, but foreshadowed
much earlier \cite{vanstockum2},
spacetimes possessing CTC
in general relativity came as a surprise to
relativists.  The shock was softened by the fact that these solutions
required unphysical stress tensor sources, and in this sense should not
have been so unexpected: it is after all a tautology that {\it any}
spacetime is the solution of the Einstein equations with {\it some}
stress tensor, as often emphasized by Synge.  Indeed, Einstein himself,
while elaborating general relativity, apparently worried about
geometries with loss of causality and CTC, and hoped that they would
be excluded by physically acceptable sources.\footnote{We
thank John Stachel for telling us this.}  Almost two decades
after G{\"o}del's work, Misner in his pioneering studies \cite{nut3}
of NUT space
showed how CTC could be generated as global effects, by taking local
expressions for a metric and making appropriate identifications among
the points.

Very recently, the subject of CTC was revived in two quite different
contexts.  The first, which we shall not discuss, involves tunneling
through wormholes in $D$=4 gravity.  The second, which will be our
subject here, concerns solutions to $D$=3 Einstein theory
with point sources or equivalently, $D$=4 gravity with
infinite parallel cosmic strings
(since the latter system is cylindrically symmetric).
We will therefore
operate entirely in the reduced dimensionality.  The dramatic
simplification at $D$=3 is that the Einstein and Riemann tensors are
equivalent, so spacetime is locally flat wherever sources are absent;
consequently there is neither gravitational radiation
nor any Newtonian force between
particles.   For these reasons, the sign of the Einstein constant is not
physically determined, unlike in $D$=4. We will adopt the usual sign
here, but mention the opposite sign, ``ghost" Einstein theory, at the end.
Local flatness in $D$=3  means that all properties
are encoded in the  global structure, {\it i.e.}, in the way the
locally  flat patches are sewn together.   Indeed, we will use this
geometric approach to analyze the CTC problem.  However, for orientation we
will
begin with a brief discussion in terms of the analytic form of the metric.

Consider the general solution outside a localized physical source
as given by the ``Kerr'' metric  \cite{djt4}
\renewcommand{\theequation}{1.\arabic{equation}}
\setcounter{equation}{0}
\begin{equation}
ds^2 = -d (t + J\th )^2 +dr^2 + \a^2 r^2 d\th^2  \; .
\end{equation}
Our units are $\k^2 = c = 1$, and   $\a \equiv 1-{M/ 2\p}$.
The constants of motion are the energy
$M$ and angular momentum  $J$
(space translations not being well defined \cite{henneaux5}.   This
interval is manifestly locally flat in terms of the redefined coordinates
$\TH = \a \th$, $T = t + J\th$,
but the global content lies in the different range,
$0 \leq \TH \leq 2\pi \a$,  of $\TH$ corresponding to the usual
conical identification,
and in the time-helical structure resulting from the fact
that the two times $T$
and $T+2\pi J$ are to be identified whenever a closed spatial circuit is
completed. The interval (1.1)
can clearly support CTC; for example, the interval traced by a circle at
constant $r$ and $t$,
\begin{equation}
\D s^2 = (2\pi \a )^2 (r^2 - J^2 /\a^2 ) < 0 \; ,
\end{equation}
is timelike for $r < |J|/\a .$
However, the relevant physical question is whether the
constituent particles
are ever confined within this radius; otherwise the CTC criterion (1.2)
ceases to apply.
To be sure, if we simply insert the metric (1.1) into the Einstein
equations, it is valid down to $r = 0$; the ``source" is a spinning particle
with $T^0_0 \sim m  \d^2 (r)$, $T^i_0 \sim J \, \e^{ij} \pa_j \, \d^2
(r)$.
But we do not accept classical spinning particles as physical,
precisely because of their singular stress tensors, any more than we do
G\"{o}del's sources.  Instead, one must check whether a system of
moving spinless particles with {\it orbital} angular momentum
can support CTC.  This was the question that was
initially  raised in \cite{djt4}
and answered in the negative, on the simple physical grounds that the
point
particles, being essentially free, will --- both initially and finally ---
be dispersed so that the constant $J$ would have been exceeded at
$t = \pm \infty$ by the radius at which the exterior metric (1.1) is
valid.  Thus, CTC, if present at all, would have to
appear and then disappear spontaneously in time in an
otherwise normal Cauchy evolution, and it seemed unlikely that
this violation of Cauchy causality would occur in a finite time region for
an otherwise non-pathological system.

It was therefore quite surprising when, a year ago, Gott \cite{gott6}
gave
an explicit construction of  a geometry generated by an
apparently acceptable source consisting
of two massive particles passing by each other at
subluminal velocities,  in which CTC appear only during
a limited time interval \cite{ori7}.  However, it was then shown both
that, in these spaces CTC will also be present at spatial infinity,
which constitutes an unphysical boundary condition, and that
the spacetimes have an imaginary (tachyonic) total  mass
\cite{djtctc8}.  This is
in contrast to the globally flat space of special relativity,
where a collection of subluminal particles cannot of course be tachyonic.
Indeed, the fact that in $D$=3 everything
lies in the global  properties raises a cautionary, and as we shall see,
decisive, note.  Let us illustrate this with one simple object
lesson for the case of static sources.  It is clear from (1.1) that a
single particle cannot have a mass greater than $2\p$ (in our units); indeed,
$m=2\p$ corresponds to a cylindrical rather than conical 2-space.  One
might suppose that, since there are no interactions in $2+1$ dimensions,
two stationary particles should give rise to a perfectly well-defined metric as
long as
each one separately satisfies the above inequality. This in fact is
not so; the sum of the two masses
must also not exceed $2\pi$.  If it does, the total mass must then
jump to the value $4\pi$ and at least one further particle is required
to be present, the total system now having an $S_2$ --- rather than an open ---
topology:
$G^0_0$ is essentially the Euler
density of the 2-space \cite{djt4}.  This example reflects
the presence of  effective global constraints in $2+1$
dimensions, even though the theory is locally trivial, so
that a source distribution consisting of several
individually acceptable particles is {\it not} thereby guaranteed to be
itself physically acceptable. The moral applies to the Gott
pair and, as we shall see, also to its lightlike extension.
We emphasize that the pathology here is not merely that there
is a total spacelike momentum, but more importantly, that the latter
implies a ``boost-identified"  exterior geometry, namely one in which
CTC will always be present at spatial infinity.  But
if one allows pathology in the boundary
conditions of any system, then it is no surprise that it will be present
in the interior as well!  Indeed, this is just the sort of behavior that the
Misner identifications \cite{nut3} gave, and can be seen in the metric
form of
the interval as well.  For, to say that the effective source is a
tachyon, really means that the exterior geometry is that generated by
an effective pointlike stress-tensor which replaces the
$T^{00} \sim \d (x)\d (y)$ of a particle with a
$T^{yy} \sim \d (x)\d (t)$, etc.  Consequently, in (the cartesian
coordinate form of) the Kerr line element (1.1), the $(x,y)$ space is
replaced by $(x,t)$ with a jump in $t$ replaced by one in $y$
\cite{djtctc8}.
The resulting metric shows that CTC do not really appear and
disappear spontaneously in some finite region where the particles pass
each other, but rather that they are always present at spatial infinity;
thence they close in (very rapidly!) on the finite interaction region.

Attempts to remedy these difficulties by adding more particles
\cite{kabat9} to the system fail; it
has been shown that the total momentum of any system containing  the
Gott time machine is necessarily tachyonic \cite{cfgb10}; thus, two
particles
 constituting the  Gott system cannot arise from the decay of a pair of
static particles (of allowed mass less than $2\pi$) since the latter's
momentum is timelike \cite{cfga11}.
If the mass of the initial static particles exceeds $2\pi$,
the universe closes; but as was shown in \cite{thooft12},
a closed universe will end in a big crunch just before the
CTC appear. Since there is no spatial infinity, the pathology there has
been transmuted into a singularity!

In this paper, we extend the Gott construction to systems involving
lightlike particles (``photons''). Here too, CTC will appear just as the the
system becomes tachyonic.  In Section 2, we review
the geometries due to a two-photon source and to the ``mixed" system
consisting of one photon and one massive
particle \cite{ds13}. These systems will be
our testing ground for the existence of CTC. In Section 3,
 we calculate the mass
for these two-particle systems, thereby
 obtaining the condition for their total momentum
to be non-tachyonic. In Section 4, the condition for CTC to arise is
derived and is shown to coincide exactly with the condition that the
system be tachyonic. Furthermore, it will be manifest that
(since they first occur there)
CTC exist at spatial infinity if they are present at all.
In Section 5, the analysis is extended to a more general model,
topologically massive gravity.
Its two-photon solution is constructed and is shown to exclude
CTC for all positive values of the photons' energies. This is true for the
ghost Einstein theory as well.

\section{Spacetimes Generated by Lightlike Sources}

In this section, we review two systems, involving lightlike sources, from
which we will attempt to build a time machine. The first consists of two
non-colliding photons, the second of one photon and one
massive particle.  Each can be obtained by pasting together the
appropriate one-particle solutions, which we first describe.

In $D$=3, a vacuum spacetime is specified by the way in which
locally flat patches are sewn together.  Different patches are identified
using
Poincar\'{e} transformations, since these define the symmetries of flat
space.  This method of constructing solutions, in which the particle
parameters
(mass, velocity, and location) determine the transformation generators,
was
presented\footnote{Such procedures are
described more formally in \cite{arizona14}.}
in \cite{djt4}.  This procedure is, of
course, completely  equivalent to the standard analytic
approach of obtaining the  metric from the field equations.

The simplest example is the conical spacetime describing
a particle of mass $m$ at rest at the origin of
the $x\!-\!y$ plane.  This solution is obtained
 by excising  a wedge of
angle $m$ with vertex at the origin and identifying the two edges
according to $x^{\pr} = \O_m x$, where  $\O_m$ is a rotation by $m$
\renewcommand{\theequation}{2.\arabic{equation}}
\setcounter{equation}{0}
\begin{equation}
\Omega_m =
\left(
\begin{array}{ccc}
 1 & 0  & 0\\
 0 & \cos\,m & \sin\,m \\
 0 & - \sin\,m & \cos\,m
\end{array}
\right) \; ,
\end{equation}
whose rows and columns are labelled by $(t,x,y)$.  This description is
completely equivalent to the metric form (1.1) with $J$=0.

The geometric description of the one-photon solution
can be found from the analytic solution, or by an
infinite boost of the conical static metric \cite{ds13}.
Consider a single photon moving along the $x$-axis with energy $E$
and energy-momentum tensor $T_{\m\n} = E\d(u) \d(y) l_{\m}l_{\n},\,
l_{\m} = \pa_{\m} u$ where $u=t-x$,
$v=t+x$ are the usual lightcone coordinates. The
Einstein equations $G_{\m \n} = T_{\m\n}$ can be
solved with a plane-wave ansatz
\begin{equation}
ds^2= ds_0^2 + F(u,y)du^2 \; ,
\end{equation}
where $ds_0^2 =  - dudv + dy^2 $ is the flat metric.
This ansatz simplifies the Einstein tensor  to
$G_{\m\n} = -\frac{1}{2} \frac{\pa^2 F}{\pa y^2}\,l_{\m} l_{\n}$,
and reduces the Einstein equations to the ordinary differential equation
$\frac{\pa^2 F}{\pa y^2} = -2 E \d(u) \d(y)$.
Solving for $F$ yields the general one-photon solution:
\begin{equation}
ds^2 = ds_0^2  -2Ey\theta (y) \d (u)
du^2
\end{equation}
up to a homogeneous solution, of the form $F = B(u)y + C(u)$,
that can be absorbed by a coordinate transformation.

If we now apply the coordinate transformation
$v\ra v  - 2 E y\theta (y) \theta (u),$
the metric becomes
\begin{equation}
ds^2  =  \theta (u)  \{ -du d ( v - 2E y\theta (y))
  + dy^2  \}
 +  \theta (-u)  \{ -dud v + dy^2 \}\; .
\end{equation}
In this form, the geometric description of the one-photon solution
becomes clear. It corresponds to making a cut along the $u=0,\,y>0$
halfplane extending
from the  photon's worldline to infinity and then identifying $v$
on the $u=0^-$ side with $v-2Ey$ on the $u=0^+$ side. It is easily
checked that the points being identified are
in fact related by the Lorentz transformation
\begin{equation}
N_E  =
\left(
\begin{array}{ccc}
1 + \textstyle{\frac{1}{2}}E^2 & - \textstyle{\frac{1}{2}} E^2  & E \\
\textstyle{\frac{1}{2}} E^2  & 1 - \textstyle{\frac{1}{2}} E^2  & E \\
E & -E & 1
\end{array}
\right) \; .
\end{equation}
This matrix corresponds to the $\b\ra 1,\, m\ra 0$ fixed energy
$E = \frac{m}{\sqrt {1-\b^2}}$,
limit of the
boost-conjugated rotation matrix $\L_\beta\O_m\L_\beta^{-1}$.
Here $\L_{ \b}$
is a Lorentz  boost in the $x$-direction,
\begin{equation} 
\L_{\b} = \frac{1}{\sqrt{1-\b^2}}
\left(
\begin{array}{ccc}
1 &  \b & 0 \\
\b&   1 & 0 \\
0 &   0 & 1
\end{array}
\right) \; .
\end{equation}
[This geometric construction of the one-photon solution is
analogous to that of the Aichelburg--Sexl one-photon geometry \cite{as15}
in  $D$=4, our null boost being the analog of their null shift.]  The above
formulation is not
unique however; an equivalent one, more analogous to the conical solution, is
obtained by boosting the cone
along its bisector rather than perpendicular to it \cite{thooftpc16}.
The physics is of course independent of such choices.

The solution for two non-colliding photons
can now be constructed by pasting together the individual one-photon
solutions.  [It is of course not possible to construct the two-photon
solution in this way
in $D$=4, since spacetime is not flat between sources.]
We consider two non-parallel\footnote{The solution for
parallel photons can also be constructed, but  it does not admit
CTC for the same reason as that given below for the one-photon
solution.} photons in their  center-of-momentum  frame,
where the photons are taken to be moving with energy $E$ respectively
in the positive $x$-direction along  $y={a},\; (a>0)$
and in the negative $x$-direction along  $y=-{a}.$
The spacetime associated with the first photon is obtained by making
a cut along the $u=t-x=0,\,y>{a}$ halfplane and then
identifying the point $(x,x,y)$ on the $u=0^-$ side with the point
$(x-E(y-{a}), x-E(y-{a}), y)$ on the $u=0^+$ side. For the second photon,
one makes a cut along $v=t+x=0,\,
y<-{a}$ and identifies $(-x,x,y)$ on $v=0^-$ with $(-x+E(y+{a}),
(x-E(y+{a}), y)$ on $v=0^+$. The complete two-photon geometry then
consists of these two one-photon solutions
simply pasted together along the $y=0$ plane.  In contrast to the
massive Gott pair, no relative boost between the two
particles' halfspaces is necessary, since the photons' motion
is already encoded in the identification made on their respective
halfplanes.

The second system, consisting of a photon and a massive particle, can
also be obtained by pasting together  the respective one-particle
solutions.  We use a frame in which the massive particle is
at rest at the origin in the $x\!-\!y$ plane, and the photon is moving in the
positive $x-$direction along $y=a>0$. For the static massive particle we
excise from the $y<a/2$ halfspace a
wedge of angle $m$  whose vertex is at the origin and which is oriented in
the negative $y$-direction, then identify the edges as usual.
For the photon, we simply
translate the one-photon solution given above
from $y=0$ to $y=a$.  If $m<\pi$, then the orientation of the wedge
ensures that there is no intersection with the photon's halfplane, so that the
two-particle solution is obtained by pasting
together these two halfspaces along $y={a/ 2}$, again with no relative
boost.  If $m>\pi$, the solution is no longer obtainable by simple
gluing, as the tails of the two sources would now overlap.

\section{Total Energy and Tachyon Conditions}

In this section, we calculate the total mass of the two
previously described systems.  In the following section,
we will see that CTC arise precisely when the total mass becomes
imaginary, {\it i.e.},
the system becomes tachyonic and
non-physical Misner identifications emerge. The mass can be found by
composing  the one-particle identifications and  writing  the result for
the complete system in the
form of  the general spacetime identification
\renewcommand{\theequation}{3.\arabic{equation}}
\setcounter{equation}{0}
\begin{equation} 
x^{\prime}  = a + {\cal L}  ( x - a) +  b \; .
\end{equation}
The spatial vector $a = (0,{\bf a})$ describes the location of the
center-of-mass;  the direction and magnitude of the timelike vector
$b$ define, respectively, the time axis and the time-shift
along it. For a
system of total mass $M$ and velocity $ \b $ in the $x$ direction,
${\cal L}$ will be a Lorentz transformation of the form
$ {\cal L} =\L_{ \b} \Omega_M \L_{\b}^{-1}$, implying in particular
that
\begin{equation}
\cos  M = \textstyle{\frac{1}{2}} ({\rm Tr}\, {\cal L}-1) \; .
\end{equation}
The system is non-tachyonic provided $M$ is real, implying that the
right-hand side lies in the range $[-1,1]$. We now proceed to
calculate ${\cal L}$, and from it $M$,
in terms of the constituent parameters of
the systems constructed in the
previous section.

For the two-photon system, the one-particle identifications
are given by
\begin{eqnarray} 
x_1^{\prime} & =  & a + N_E ( x_1- a) \nonumber \\ [.1in]
x_2^{\prime} & =  & -a + \O_{\pi} N_E \O_{-\pi} ( x_2 + a) \; .
\end{eqnarray}
The conjugation of $N_E$ by a $\pi$-rotation in the second equation
reflects the fact that the second photon is moving in
the negative $x$-direction. Composing the two identifications in
(3.3), we find ${\cal L} = N_E\O_{\p} N_E \O_{-\p}, $
and ${\rm Tr}\, {\cal L}= 3-4E^2 +E^4$.  Comparing
with (3.2), we obtain the condition
\begin{equation}
E > E_{\max} =2
\end{equation}
for the system to be tachyonic. This condition can also be formally
obtained as the limit of the original Gott
condition \cite{gott6}
for two masses $m$ moving subluminally: there, the tachyonic
threshold is given by
\begin{equation}
\frac{\sin\textstyle{\frac{1}{2}} m}{\sqrt{1-\b^2}} > 1 \; .
\end{equation}
Clearly, the limit $m\ra 0$, $\b\ra 1$ in this equation
(with the energy fixed) yields (3.4).

For the mixed system, the one-particle identifications are given by
\begin{eqnarray}
x_1^{\pr} & = & a + N_E (x_1-a) \; , \nonumber \\[.1in]
x_2^{\pr} & = & \O_m x_2 \; .
\end{eqnarray}
Composing these yields ${\cal L} = N_E \O_m$; comparison of its trace
with (3.2)
implies
\begin{equation}
\cos  M = \cos m - (\sin  m)
E  + \textstyle{\frac{1}{4}}  (1-\cos m)  E^2 \; .
\end{equation}
The criterion for tachyonic $M$ can be expressed as a condition on $E$
for fixed $m$,
\begin{equation} 
E >  E_{\max}  =  2 \:
\frac{\sin m + \sqrt{2(1-\cos m)}}
{1-\cos m} \; .
\end{equation}

\section{Closed Timelike Curves}

We now find the conditions for CTC to be present in the two-photon and
mixed systems.  We first show that the one-photon spacetime (like the
conical spacetime for a particle of non-zero mass) does not admit CTC.
This may not be  obvious, since  the identification
involves a timeshift, which as in the case of the Kerr solution
(1.1), could potentially lead to CTC.   Recall that the one-photon
solution is characterized by the shift $(x,x,y)\ra (x^{\pr}, x^{\pr},
y^{\pr}) =  (x- Ey, x - Ey, y)$ upon crossing the $u=t-x=0,\,y>0$ null
halfplane from $u=0^-$ to $u=0^+$.  A CTC, $\g$, would have
to cross this halfplane to take advantage of the
timeshift (there being no CTC within flat spacetime patches).
Irrespective of where $\g$ enters the halfplane, the
shifted point from which $\g$ emerges is
separated from the entry point by a lightlike interval. Therefore only
particles travelling faster than the speed of light can complete the loop
in time, showing that the one-photon solution does not admit CTC.

Now consider the two-photon solution described by the identifications
(3.3).  Since, as shown
above, the one-photon solution does not permit CTC, we can restrict
ourselves to curves $\g$ that
enclose both photons' worldlines and therefore intersect both of their
halfplanes. In order that it not become spacelike,
$\g$ must be directed opposite to the photon whose plane it is about to
cross; this sense will automatically yield a  gain in time upon
crossing the halfplanes.
Label the
 point at which $\g$ intersects the $u=0$, $y>a$ halfplane
by  $x^{\m}_1 = (x_1,x_1,y_1)$ on the $u=0^-$ side and hence by
$x^{\m\pr}_1 = (x_1-E(y_1-{a}), x_1-E(y_1-{a}), y_1)$ on the $u=0^+$ side,
and the point at which $\g$ intersects the other halfplane by
$x^{\m}_2 = (-x_2, x_2, y_2)$ on the $v=0^-$ side and hence by
$x^{\m\pr}_2 = (-x_2+ E(y_2+{a}), x_2 -E (y_2+{a}), y_2)$ on the
$v=0^+$ side.  For $\g$ to be timelike, the total traversed
distance,
\renewcommand{\theequation}{4.\arabic{equation}}
\setcounter{equation}{0}
\begin{eqnarray}
d & = & |{\bf x_2} - {\bf x_1^{\pr}}| +
        |{\bf x_1} - {\bf x_2^{\pr}}| \\[.2in]
  & = & \sqrt{(x_1\!-\!x_2\!-\!E(y_1\!-\!a))^2\!+\!
(y_1\!-\!y_2)^2}
+ \sqrt{(x_1\!-\!x_2\!+\!E(y_2\!+\!a))^2\!+\!(y_1
\!-\!y_2)^2 },
\nonumber
\end{eqnarray}
must be less than the total elapsed time,
\begin{equation}
T = (t_2 -t_1^{\pr}) + (t_1 -t_2^{\pr})= E(y_1-y_2-2a) \; .
\end{equation}
For a given $T$, we can find the minimum value of $d$ as a function
of its arguments. The extremization occurs at $x_1-x_2 = T/2$ and $y_1 +y_2
=0$, with the result that $d_{\min} = 4y_1$,
with $T = 2E(y_1 -a)$. Therefore CTC will be present if
\begin{equation} 
 y_1/a > E/(E-2)  \; .
\end{equation}
Recalling that $y_1 > a > 0$, we see that the lowest allowed value is
$E=2$, precisely the tachyon threshold
$E_{\max}$ of  (3.4); there, $y_1$ (and
therefore also $-y_2$) becomes infinite, {\it i.e.,} CTC first arise at
spatial infinity.  As the energy increases, the CTC spread into the
interior as well, but they are
always present at spatial infinity, if present at all.
[The requirement that $\gamma$ be everywhere future-directed
imposes no relevant conditions:  The individual time segments
$(t_2 - t_1^\prime )$ and
$(t_1 - t_2^\prime )$ must each be positive, implying the inequalities
$E(y_1 - a) > (x_1 + x_2 ) > E(y_2 + a)$.  At the minimum, they
read $|x_1 + x_2 | < E(y_1 - a)$, and are easily
satisfied, since $(x_1 + x_2)$ is otherwise unconstrained.]

Let us now find the condition for  which the ``mixed" system admits
CTC. Since neither individual
one-particle solution alone admits CTC,  a potential CTC, $\g$, must
again enclose both particles, thereby intersecting both the photon's
halfplane and the static particle's wedge. Let the point of intersection
with the halfplane be labelled by  $x^{\m}_1
 =(x_1,x_1,y_1)$ on the $u=0^-$ side and by
$x_1^{\m\pr} = (x_1- E (y_1-a), x_1- E (y_1-a), y_1)$ on
the $u=0^+$ side, and that  with the wedge
by $x^{\m}_2 =  (t_2,y_2{\rm
tan}\;\frac{m}{2},y_2)$ on one edge and by the rotated values
$  x_2^{\m\pr} = (t_2, -y_2{\rm
tan}\;\frac{m}{2},y_2) $ on the other edge.
The curve $\g$ is timelike provided the
distance traversed,
\begin{equation}
d = \sqrt{ (x_1\!-\!E(y_1\!-\!a)\!-\!y_2\tan \,\textstyle{\frac{m}{2}}
)^2
\!+\!(y_1\!-\!y_2)^2}  +
\sqrt{(x_1\!+\!y_2\tan\,\textstyle{\frac{m}{2}} )^2
\!+\!(y_1\!-\!y_2)^2}\; ,
\end{equation}
is less than the elapsed time $T = E(y_1-a)$.
Here, we minimize $d$ with respect to $x_1$ and $y_2$ for fixed $T$
(or $y_1$); the extremum occurs at $x_1 = T/2$, $y_2 = y_1 \cos \,
\frac{m}{2} - \frac{1}{2} E
(y_1 -a ) \sin \, \frac{m}{2}$ and is given by $d_{\min} = 2a \sin\,
\frac{m}{2}  + (y_1 -a) (E {\rm cos}\, \frac{m}{2} + 2 {\rm sin}\,
\frac{m}{2})$. Therefore, existence of CTC requires
\begin{equation} 
y_1/a > E/
\left(
E-2\,\frac{\sin \textstyle{\frac{m}{2}}}
{1-\cos \textstyle{\frac{m}{2}}}
\right) \; .
\end{equation}
Since $y_1 > 0$, $E$ must equal or exceed the
threshold tachyon value $E_{\max}$ of (3.8), as is easily seen using
half-angle formulas.  Again, $y_1$ is infinite at $E = E_{\max}$, and
as the energy increases, CTC begin to move into the finite region.
[Here the requirements that the travel segments be future-directed
reduce to $E(y_1 - a) > (x_1 - t_2 ) > 0$, which can  always be fulfilled
by adjusting $t_2$.]

\section{Topologically Massive Gravity}

Topologically massive gravity (TMG)
\cite{tmgt17} is of
interest because, in contrast to pure $D$=3 gravity, it is a dynamical
theory.  In \cite{ds13} exact solutions for lightlike
sources, including (for certain orientations) two-photon
solutions,  were found.  Here we show that they do not admit
CTC for any values of the photons' energies. [We cannot
construct the analog of the ``mixed'' system since the exact solution for
a massive source is not known in TMG.] The field
 equations for TMG are  the ghost ({\it i.e.,}  with the opposite sign of
$\k^2$) Einstein equations,  to which is added the conformally
invariant, conserved, symmetric  Cotton
tensor $C^{\mu\nu} \equiv  \e^{\m\a\b} D_\a (R_\b~^\n -
\frac{1}{4} \d^\n_\b R)$:
\renewcommand{\theequation}{5.\arabic{equation}}
\setcounter{equation}{0}
\begin{equation}
E_{\mu\nu} \equiv G_{\mu\nu} + \textstyle{\frac{1}{\mu}}
C_{\mu\nu} = -\kappa^2 T_{\mu\nu}\; .
\end{equation}
Here $\mu$ is a parameter (whose sign is arbitrary) with dimensions
of mass or inverse length.  These equations
can be solved exactly for a photon source \cite{ds13}, using the
plane-wave ansatz (2.2). For a photon with energy $E$ moving
along the positive $x-$axis, the resulting spacetime metric is
given by
\begin{equation}
ds^2 = -dudv +dy^2 + 2 \k^2 Ef(y)\delta(u) du^2,
\;\;\;
f\equiv (y+ \textstyle{\frac{1}{\m}}(e^{-\m y}-1))\theta (y)\;.
\end{equation}
Observe that as $\m\ra\infty$, $f(y)\ra y\theta (y)$ and
(5.2) reduces to the ghost gravitational one-photon solution ({\it
i.e.,} (2.3) with the opposite sign
of $\k^2$).  Like its Einstein counterpart, the metric (5.2)
can also be obtained by a cut-and-paste
procedure, albeit not by using Poincar{\'e} transformations, since the
spacetime is not flat along the $u=0$, $y>0$ null halfplane.
After applying the coordinate transformation $v \rightarrow  v + 2\k^2 E f(y)
\theta (u)$ to remove the $\delta (u)$ factor in (5.2),
it takes the  form
\begin{eqnarray}
ds^2  & = & -du d v + dy^2 - 2\k^2 E  \theta(u) f^{\prime}(y)
dudy \nonumber \\[.1in]
      & = & \theta (u)  \{ -du d ( v + 2\k^2 E f(y))  + dy^2  \}
 +  \theta (-u)  \{ -dud v + dy^2 \} \; .
\end{eqnarray}
Clearly, this corresponds to identifying $v$ on $u=0^-$ with
$v + 2 \k^2 E f(y)$ on $u=0^+$.

We can construct the two-photon spacetime, in the convenient frame
where one photon moves in the
positive $x$-direction along $y={a}>0$ and the other in the negative
$x$-direction along $y=-{a}$,
 by pasting together the one-photon solutions along
the $y=0$ hyperplane.
This pasting is possible since, as in Einstein
gravity,  each of the one-particle
solutions is both flat  and has zero extrinsic curvature on this
hyperplane. The resulting spacetime consists in making
cuts  along the $u=0,\;y>{a}$ and $v=0,\;y<-{a}$
halfplanes and then identifying
the   point $(x_1,x_1,y_1)$ on the $u=0^-$ side with $ (x_1+\k^2 E
f(y_1-{a}), x_1+ \k^2 E f(y_1-{a}), y_1)$ on the $u=0^+$ side, and the
point $(-x_2,x_2,y_2)$ on the $v=0^-$ side with $ (-x_2-\k^2 E
f(-y_2-{a}), x_2 + \k^2 E f(-y_2-{a}), y_2)$  on the $v=0^+$ side.
[We note that if the directions of both photons were reversed
(corresponding to
the parity operation $x\ra -x$), then this simple
pasting prescription is not possible, since
the individual one-photon solutions are no longer
flat on the $y=0$ hyperplane. This reflects the parity violation
implicit in the dependence of the field
equations (5.1) on $\epsilon_{\m\n\r}.$]
The absence of CTC for all positive values of $E$ can be seen as follows.
Since $f(y)\ge 0$,
the time shift upon crossing a halfplane has opposite sign relative to
that of the pure gravity case,
(2.3).  Hence, to gain a timeshift one would have to cross the
(null) halfplane from
the $u>0$ side, rather than from the $u<0$ side,
which  is impossible for any timelike (or lightlike) curve.
These conclusions obviously also apply to ghost Einstein
gravity\footnote{As we mentioned at the outset, there is no
{\it a priori} physical
requirement that $\k^2$ be positive within the $D$=3 picture, except
if one wishes to regard it as a reduction  from $D$=4 in order to make
contact with cosmic strings.} as
well, since the latter is just the $\m\ra\infty$ limit of TMG.

\section{Conclusion}

We  have examined, in the case where one or both of their sources
are lightlike, the physical difficulties associated with geometries that
permit CTC in $2+1$ Einstein gravity. We first obtained the total
energy in terms of the constituent parameters,
 using the geometric approach in which flat
patches are identified through null boosts, and found
the  conditions for the systems
to be tachyonic.  We then showed that, as the energy of a system first
surpasses its tachyonic bound, CTC initially
emerge at spatial infinity, then spread into the interior, but always
remain present at infinity.
 This is, of course, the manifestation of the unphysical
spatial boundary conditions that are the price paid for CTC, and
are analogous  to
the Misner identifications that give rise to CTC in NUT space.
We also demonstrated,
using the known two-photon solution in TMG, that this
dynamical model, and therefore also its limit,
ghost Einstein theory, never admits  CTC.

All sources of the $2+1$ Einstein equations considered to date thus
share the property that if they are physical---nontachyonic---they do not
engender acausal geometries. These results add evidence in favor of
both Einstein's original hope and  its  recent avatar
\cite{hawking18}, that this is a universal property of general relativity.

\section{Acknowledgements}

We thank G. `t Hooft for useful conversations.
This work was supported by the
National Science Foundation under grant \#PHY88--04561.

\end{document}